\documentclass[12pt, preprint]{aastex}

\begin{document}

\title{The Scattering Polarization of the Sr {\sc i} 4607 \AA\ Line at the Diffraction Limit Resolution of a 1-m Telescope}

\shorttitle{The Scattering Polarization of the Sr {\sc i} 4607 \AA\ Line}

\author{Javier Trujillo Bueno\altaffilmark{1,3} and
Nataliya Shchukina\altaffilmark{2}}
\altaffiltext{1}{Instituto de Astrof\'{\i}sica de Canarias, 38205 La Laguna,
Tenerife, Spain}
\altaffiltext{2}{Main Astronomical Observatory, National Academy of Sciences, 27 Zabolotnogo Street, Kiev 03680, Ukraine}
\altaffiltext{3}{Consejo Superior de Investigaciones Cient\'{\i}ficas (Spain)} \email{jtb@iac.es, shchukin@mao.kiev.ua}

\date{{\bf Accepted for publication in The Astrophysical Journal Letters (2007)}}

\begin{abstract}

One of the greatest challenges in solar and stellar physics in coming years will be to observe the Second Solar Spectrum with a spatial resolution significantly better than 1 arcsec. This type of scattering polarization observations would probably allow us to discover hitherto unknown aspects of the Sun's hidden magnetism. Here we report on some theoretical predictions for the photospheric line of Sr {\sc i} at 4607 \AA, which we have obtained by solving the three-dimensional (3D) radiative transfer problem of scattering line polarization in a realistic hydrodynamical model of the solar photosphere. We have taken into account not only the anisotropy of the radiation field in the 3D medium and the Hanle effect of a tangled magnetic field, but also the symmetry breaking effects caused by the horizontal atmospheric inhomogeneities produced by the solar surface convection. Interestingly, the $Q/I$ and $U/I$ linear polarization signals of the emergent spectral line radiation have sizable values and fluctuations, even at the very center of the solar disk where we meet the forward scattering case. The ensuing small-scale patterns in $Q/I$ and $U/I$ turn out to be sensitive to the assumed magnetic field model, and are of great diagnostic value. We argue that it should be possible to observe them with the help of a 1-m telescope equipped with adaptive optics and a suitable polarimeter.

\end{abstract}

\keywords{Sun: atmosphere; polarization; scattering; radiative transfer; Sun: magnetic fields; stars: magnetic fields}

\section{Introduction}

The observation and modeling of the spectral line polarization produced by radiatively induced quantum coherences in atomic systems provides a novel diagnostic window to explore the magnetism of the Sun and of other stars via the Hanle effect (e.g., the recent monograph by Landi Degl'Innocenti \& Landolfi 2004). For the moment, such scattering polarization signals have been observed without or with poor spatial and/or temporal resolution (e.g., Stenflo \& Keller 1997; Trujillo Bueno et al. 2001; Malherbe et al. 2007). However, as shown in this work, there are important scientific reasons to try to observe them with a spatial resolution significantly better than 1 arcsec, so as to be able to see clearly the details of the solar granulation pattern.

In the apparently quiet regions of the solar atmosphere there are three main reasons for expecting spatial variability in the linear polarization signals produced by scattering processes. First, the atmospheric inhomogeneities produced by the solar surface convection lead to horizontal fluctuations in the anisotropy of the radiation field, which must have their impact on the polarization of the emergent radiation (Trujillo Bueno 2003; Trujillo Bueno et al. 2004). Second, the same atmospheric inhomogeneities should lead to local symmetry breaking effects, so that the radiation field loses the axial symmetry that is characteristic of a plane-parallel or spherically symmetric stellar atmosphere model. As a result, as pointed out by Manso Sainz \& Trujillo Bueno (1999) through two-dimensional radiative transfer calculations in simplified solar model atmospheres, one can have scattering line polarization even at the center of the solar disk, without the need of an inclined magnetic field. Third, the scattering polarization signals are sensitive to magnetic fields via the Hanle effect, and could therefore show fluctuations across the observed field of view if the statistical properties of the hidden field are not homogeneous, as is indeed indicated by a joint analysis of the Hanle effect in atomic and molecular lines (Trujillo Bueno et al. 2004; 2006). 

In this Letter we present simulated polarimetric observations for the well-known case of the Sr {\sc i} 4607 \AA\ line, which we have obtained by solving the radiative transfer problem of scattering line polarization in a realistic three-dimensional (3D) hydrodynamical model of solar surface convection. Observations of the spatial variations of the scattering polarization in the Sr {\sc i} 4607 \AA\ line have shown that these variations are very small and next to invisible with the used resolution (e.g., Stenflo et al. 1997; Trujillo Bueno et al. 2001; Malherbe et al. 2007), in contrast to the large spatial variations in $Q/I$ and $U/I$ seen in much stronger lines like Ca {\sc i} 4227\,\AA\ (e.g., Fig. 5 in Stenflo 2004) and Ca {\sc ii} K (e.g., Fig. 4 in Stenflo 2006), which have been tentatively interpreted in terms of largely resolved magnetic fields in the solar chromosphere. While the chromosphere has prominent inhomogeneities on the supergranular scales, which are well resolved in the mentioned observations, the quiet photosphere varies on the scale of the solar granulation, which has so far been below the resolution that has been achieved in combination with high-precision spectro-polarimetry. 

In our previous 3D radiative transfer investigations (Shchukina \& Trujillo Bueno 2003; Trujillo Bueno et al. 2004) only the influence of the mean intensity and anisotropy of the Sr {\sc i} 4607 \AA\ line radiation was accounted for. Here we consider the full 3D scattering line polarization problem in the same realistic hydrodynamical model of the solar photosphere, including the Hanle effect of a microturbulent magnetic field, but taking into account the above-mentioned symmetry breaking effects that we had previously neglected. As we shall see, we confirm our conclusions concerning the presence of a substantial amount of hidden magnetic energy in the quiet regions of the solar photosphere. In addition, we report on the new interesting results summarized in the abstract.

\section{Formulation of the problem}

We have solved the 3D scattering polarization problem for the Sr {\sc i} 4607 \AA\ line in a realistic model of the solar photosphere resulting from the hydrodynamical simulations of solar surface convection by Asplund et al. (2000). 

First, we calculated the number density of Sr {\sc i} atoms at each grid-point of the 3D photospheric model and the overall population of the lower and upper levels of the 4607 \AA\ line transition, whose total angular momentum values are $J_l=0$ and $J_u=1$, respectively. To this end, we solved the standard non-LTE radiative transfer problem including all the allowed radiative and collisional transitions between the 15 bound levels of a realistic model of Sr {\sc i} and the ensuing ionizing transitions to the ground level of Sr {\sc ii}. It is important to calculate the ionization balance of strontium without assuming LTE 
and using a realistic atomic model in order to account properly for the UV overionization mechanism by means of which all the Sr {\sc i} levels become significantly underpopulated with respect to the LTE case (Shchukina \& Trujillo Bueno 2003). We point out that with the resulting self-consistently calculated populations, the computed spatialy-averaged emergent intensity profiles (which take into account the Doppler shifts of the convective flow velocities in the 3D model) turn out to be in excellent agreement with the observations.

Our second step consisted in calculating, at each grid-point of the 3D model, the self-consistent values of the six multipolar components of the atomic density-matrix that characterize the excitation state of the upper level of the Sr {\sc i} 4607 \AA\ line transition, using the previously calculated lower level population values. Such density-matrix elements quantify the overall population of the upper level (via the $\rho^0_0$ component), the level's population imbalances (via the $\rho^2_0$ component) and the quantum coherences between each pair of magnetic sublevels pertaining to the upper level (via the real and imaginary parts of the $\rho^2_1$ and $\rho^2_2$ components).
To this end, we solved iteratively the set of equations that results from combining the radiative transfer equations for the Stokes parameters $I$, $Q$ and $U$ and the following statistical equilibrium equations, whose derivation can be found in Appendix A of Trujillo Bueno \& Manso Sainz (1999):

\begin{equation}
S^0_0\,=\,(1-\epsilon){{J}}_0^0\,+\,{\epsilon}\,B_{\nu},
\end{equation}
and

\begin{eqnarray}
  [1+{\delta}^{(2)}(1-\epsilon)]
  \left( \begin{array}{l} 
      {S^2_0} \\ 
      {\tilde{S}^2_1} \\
      {\hat{S}^2_1} \\ 
      {\tilde{S}^2_2} \\ 
      {\hat{S}^2_2}
    \end{array} \right) =  
  (1-\epsilon) \, w^{(2)}_{J_uJ_l} \, {\cal {H}}^{(2)}
  \left( \begin{array}{r}
      {J}^2_0 \\
      \tilde{J}{}^{2}_{1} \\
      -\hat{J}{}^{2}_{1} \\
      \tilde{J}{}^{2}_{2} \\
      -\hat{J}{}^{2}_{2}
    \end{array} \right)\,.
\end{eqnarray}
In these expressions, which are valid for the case of a resonance line without lower-level polarization, $B_{\nu}$ is the Planck function, ${S_Q^K}\,=\,{\frac{2h{\nu}^3}{c^2}}{\frac{2{J}_l+1}{\sqrt{2{J}_u+1}}}{\rho}_Q^K$ (with ${\rho}^K_Q$ 
the density-matrix elements of the upper level normalized
to the overall population of the ground level) and ${{J}}^K_Q$ are 
the spherical components of the radiation field tensor (e.g., \S~5.11 in Landi Degl'Innocenti \& Landolfi 2004). We point out that ${S_Q^2}$ and ${{J}}^2_Q$ (with $Q=1,2$) are complex quantities and that in Eq. (2) ${\tilde{S}^2_Q}$ and $\tilde{J}{}^{2}_{Q}$ indicate the {\em real} parts, while ${\hat{S}^2_Q}$ and $\hat{J}{}^{2}_{Q}$ the {\em imaginary} parts. The expressions we have used for ${{J}}^0_0$, ${{J}}^2_0$ and ${{J}}^2_Q$ (with $Q=1,2$) are those given by Eqs. (4)-(7) in Trujillo Bueno (2001), taking into account the Doppler shifts caused by the convective velocities. Note also that ${\cal {H}}^{(2)}$ is the Hanle depolarization factor of a microturbulent magnetic field (see Eq. A.16 in Trujillo Bueno \& Manso Sainz 1999), $\epsilon$ is the collisional destruction probability due to inelastic collisions with electrons (e.g., Shchukina \& Trujillo Bueno 2001), ${\delta}^{(2)}=D^{(2)}/A_{ul}$ is the collisional depolarizing rate due to elastic collisions with neutral hydrogen atoms measured in units of the Einstein A$_{ul}$ coefficient (e.g., Faurobert-Scholl et al. 1995), and $w^{(2)}_{J_uJ_l}$ is a coefficient whose value is unity for the Sr {\sc i} 4607 \AA\ line under consideration. 

The spherical components of the radiation field tensor are given by angular and frequency weighted averages of the Stokes parameters.  Thus, ${J}{}^{0}_{0}$ quantifies the mean intensity of the spectral line radiation, ${J}{}^{2}_{0}$ the anisotropy factor, and ${J}{}^{2}_{Q}$ (with $Q=1,2$) the breaking of the axial symmetry of the radiation field through the complex azimuthal exponentials that appear inside the angular integrals (i.e., $e^{i\chi}$ for ${J}{}^{2}_{1}$ and $e^{i2\chi}$ for ${J}{}^{2}_{2}$, where $\chi$ is the azimuth of the ray).
We point out that while ${J}{}^{2}_{1}$ and ${J}{}^{2}_{2}$ are zero in a plane-parallel or spherically symmetric model atmosphere in the absence or in the presence of a microturbulent magnetic field, they turn out to fluctuate horizontally at each height in the above-mentioned 3D hydrodynamical model of the solar photosphere, with sizable positive and negative values mainly around the boundaries between the granular and intergranular regions.
 
The statistical equilibrium equations 
have a clear physical meaning. In particular, Eqs. (2) describe the transfer of the symmetry properties of the radiation field directly to the atomic system. 
Therefore, without the need of a magnetic field (i.e., for ${\cal {H}}^{(2)}=1$) optical pumping processes can
generate both {\em population imbalances} (if ${J}^2_0{\ne}0$) and
{\em coherences} (if ${J}^2_1{\ne}0$ and/or ${J}^2_2{\ne}0$).
In the presence of a microturbulent magnetic 
field the population imbalances
($S^2_0$) and the quantum coherences ($S^2_1$ and $S^2_2$) are reduced because ${\cal {H}}^{(2)}$ varies
between 1 (for $B=0$ G) and 0.2 (for $B{>}10B_c$, where 
$B_c\,{\approx}\,(1+{\delta}^{(2)})\,1.137{\times}10^{-7}A_{ul}/g_{u}$, with $g_{u}$ the Land\'e factor of the upper level). 

Our numerical solution of this physical problem is based on the iterative methods developed by Trujillo Bueno \& Manso Sainz (1999), which require calculating the radiation field tensors ${J}^K_Q$ at each iterative step. Since the lower-level of the Sr {\sc i} 4607 \AA\ line has $J_l=0$, it cannot be polarized. Therefore, the transfer equation for Stokes $X$ (with $X=I,Q,U$) is simply given by $dX/d{\tau}=X-S_X$, which we have solved by applying the 3D formal solver of Fabiani Bendicho \& Trujillo Bueno (1999). The line contributions to $S_Q$ 
and $S_U$ are the following (e.g., Manso Sainz \& Trujillo Bueno 1999)

\begin{eqnarray}
  {S}^{line}_Q=w^{(2)}_{J_uJ_l}\Big{\{}\frac{3}{2\sqrt{2}}(\mu^2-1) 
  {S}^2_0 \nonumber
\end{eqnarray}
\vspace{-0.1in}
\begin{eqnarray}
\hspace{0.5in} -
  \sqrt{3}  \mu \sqrt{1-\mu^2} (\cos \chi
  {{\tilde S}}^2_1 - \sin 
  \chi {{\hat S}}^2_1) \nonumber 
\end{eqnarray}
\vspace{-0.2in}
\begin{eqnarray}
\hspace{0.5in} - \frac{\sqrt{3}}{2} (1+\mu^2) (\cos
  2\chi \, {{\tilde S}}^2_2-\sin 2\chi \, {{\hat S}}^2_2) \Big{\}},
\end{eqnarray}
and

\begin{eqnarray}
S^{line}_U=w^{(2)}_{J_uJ_l}\sqrt{3} \,\Big{\{} \sqrt{1-\mu^2} ( \sin \chi
  {{\tilde S}}^2_1+\cos \chi {{\hat S}}^2_1) \nonumber 
\end{eqnarray}
\vspace{-0.2in}
\begin{eqnarray}
\hspace{1.2in}  +
  \mu (\sin 2\chi \, {{\tilde S}}^2_2 + \cos 2\chi \, {{\hat S}}^2_2) \Big{\}},
\end{eqnarray}
where the orientation of the ray is specified by
$\mu={\rm cos}\theta$ (with $\theta$ the polar angle) and by the azimuthal angle $\chi$. The exact expression for ${S}^{line}_I$ that we have used in our calculations is as complicated as the two previous ones (see Eq. 14 in Manso Sainz \& Trujillo Bueno 1999), but in a weakly anisotropic medium like the solar photosphere ${S}^{line}_I{\,}{\approx}{\,}S^0_0$. Note also that at the line center of a significantly strong spectral line an approximate expression for estimating the emergent fractional linear polarization is $X/I{\approx}S^{line}_X/{S}^{line}_I$ (for $X=Q,U$), with the corresponding source function values calculated at the atmospheric height where the line center optical depth is unity along the line of sight.

It is important to point out that the above-mentioned symmetry breaking effects imply non-zero values for ${{J}^2_1}$ and ${{J}^2_2}$, which in turn imply non-zero values for ${{S}^2_1}$ and ${{S}^2_2}$ (see Eq. 2). Note also from Eqs. (3) and (4) that their main observable effects would be non-zero Stokes $Q/I$ and $U/I$ signals at the solar disk center ($\mu=1$) and non-zero Stokes $U/I$ signals at any off-disk-center position. Let us show now how large such fractional polarization signals are expected to be at three on-disk positions ($\mu=0.1$, $\mu=0.5$ and $\mu=1$), taking into account the diffraction limit effect of a 1-m telescope (which we have accounted for through convolution with the Airy function).

\section{The effects of symmetry breaking on the scattering line polarization}

Figure 1 shows the center-to-limb variation of the $Q/I$ and $U/I$ line-center signals of the Sr {\sc i} 4607 \AA\ line, which we have obtained by solving the scattering line polarization problem in the above-mentioned 3D hydrodynamical model without magnetic fields. We point out that in each panel of Fig. 1 there are some points of the field of view with signals outside the minimum and maximum values that we have chosen to optimize the visualization.

As expected, at $\mu=0.1$ and $\mu=0.5$ (see the left and middle top panels of Fig. 1) the $Q/I$ signals are almost everywhere positive, because far away from the solar disk center the term of Eq. (3) proportional to ${{S}^2_0}$ makes the dominant contribution\footnote{Note that this term is proportional to ${{J}^2_0}$, which quantifies the anisotropy factor of the radiation field. In any case, it is important to point out that the other terms of Eq. (3) (i.e., those caused by the symmetry breaking effects quantified by the ${{J}^2_1}$ and ${{J}^2_2}$ tensors) do influence the local values of $Q/I$.}. In the absence of magnetic fields $Q/I$ at $\mu=0.1$ varies between about $1\%$ and $4\%$, while the range of variation at $\mu=0.5$ lies between $-0.14\%$ and $1.55\%$. The spatially averaged 
$Q/I$ amplitude is about $2.5\%$ at $\mu=0.1$ and $0.5\%$ at $\mu=0.5$, which is in agreement with the results of Trujillo Bueno et al. (2004). The standard deviations ($\sigma$) of the $Q/I$ fluctuations are approximately $0.5\%$ at $\mu=0.1$ and $0.3\%$ at $\mu=0.5$.
As shown in the corresponding bottom panels of Fig. 1   
the Stokes $U/I$ signals are very significant, with a typical spatial scale of the fluctuation similar to that of $Q/I$, but with positive and negative values lying between about $-1\%$ and $1\%$ at $\mu=0.5$ (with a $\sigma{\approx}0.3\%$)
and between $-2\%$ and $2\%$ at $\mu=0.1$ (with a $\sigma{\approx}0.7\%$). Such $U/I$ signals are exclusively due to the symmetry breaking effects caused by the horizontal atmospheric inhomogeneities, which are quantified by the tensors ${J}^2_1$ and ${J}^2_2$. The spatially averaged $U/I$ amplitudes are not zero, although they are rather small (e.g., $\langle U/I \rangle\,{\approx}-0.03\%$ at $\mu=0.5$). 

The right panels of Fig. 1 show that the $Q/I$ and $U/I$ signals at the solar disk center ($\mu=1$) are significant and that they have subgranular patterns. In this forward scattering geometry both $Q/I$ and $U/I$ have positive and negative values, which are exclusively due to the symmetry breaking effects (see Eqs. 3 and 4 for $\mu=1$). In the unmagnetized case of Fig. 1 such values vary between $-0.6\%$ and $0.8\%$, approximately, but with most of the $Q/I$ signals located between $-0.2\%$ and $0.2\%$
(with a $\sigma{\approx}0.1\%$). As expected, the spatially averaged $Q/I$ and $U/I$ values at the solar disk center are very small --that is, of the order of $0.001\%$.

Figure 2 shows again results for the $\mu=0.5$ case, but assuming a particularly interesting magnetized model characterized by a horizontally fluctuating microturbulent field with $B=15$ G at all heights within the solid-line contours of the right panels of Fig. 1 (which delineate the granular upflowing regions) and 
$B=300$ G at all heights outside such solid-line contours (which  correspond to the downflowing intergranular regions). This model implies saturation of the Hanle effect for the Sr {\sc i} 4607 \AA\ line in the intergranular regions of the photospheric plasma, as suggested by the Hanle-effect investigation of Trujillo Bueno et al. (2004; 2006). The spatially averaged $Q/I$ amplitude is now about $0.3\%$ (instead of the $0.5\%$ value corresponding to the unmagnetized case of Fig. 1), while the amplitude of the spatially averaged $U/I$ profile is only $-0.017\%$ (instead of $-0.03\%$). 
The most remarkable point is found when comparing the patterns of Fig. 2 with those of the middle panels of Fig. 1. Now, in the magnetized case under consideration, the contrast of the $Q/I$ and $U/I$ fluctuations across the field of view 
is significantly reduced, as a result of having assumed the largest possible Hanle depolarization in the intergranular regions. A similar conclusion is obtained for the other line-of-sights considered in Fig. 1.

The use of a 1-m telescope equipped with adaptive optics should be sufficient to measure with a spatial resolution of 100 km and a temporal resolution shorter than 60 sec
the predicted linear polarization signals with a polarimetric sensitivity better than $0.1\%$. A suitable instrument for detecting the reported small-scale fluctuations in $Q/I$ and $U/I$ would be a Fabry-Perot interferometer combined with a modern polarimeter. Particular attention should be given to the spectral resolution of the instrument, since an instrumental profile with a too large full-width at half-maximum (FWHM) value would significantly reduce the amplitudes of the measured $Q/I$ and $U/I$ fluctuations. A FWHM\,${\lesssim}\,25$ m\AA\ seems to be appropiate, given that for 25 m\AA\ the minimum and maximum $Q/I$ values at $\mu=0.5$ would be $-0.08\%$ and $0.9\%$, instead of the $-0.14\%$ and $1.55\%$ values corresponding to the infinite spectral resolution case of Fig. 1. Therefore, a Fabry-Perot system based on two or three etalons would be advisable.

\section{Concluding comments}

Probably, the most interesting conclusion of this investigation is that
the horizontal atmospheric inhomogeneities caused by the solar surface convection give rise to measurable linear polarization in the emergent radiation of the Sr {\sc i} 4607 \AA\ line at the solar disk center. Such $Q/I$ and $U/I$ disk-center signals have positive and negative values fluctuating horizontally with a sub-granular pattern. They are exclusively due to the breaking of the axial symmetry of the radiation field that illuminates the atomic system, which induces quantum coherences between the magnetic sublevels of the upper level of the Sr {\sc i} 4607 \AA\ line without the need of an inclined magnetic field.

\vspace{0.2truecm} 

Actually, such symmetry breaking effects produce sizable $U/I$ signals at any on-disk position, with positive and negative values fluctuating across the field of view with an amplitude that increases towards the solar limb. On the other hand, with the exception of the previously mentioned forward scattering disk-center case, the $Q/I$ signals far away from the solar disk center are instead dominated by the anisotropy of the radiation field and fluctuate around a non-zero mean value which is the larger the closer to the solar limb. The local $Q/I$ values are influenced by the symmetry breaking effects, but the spatially averaged $Q/I$ amplitudes at each $\mu$ position turn out to be similar to those calculated by Trujillo Bueno et al. (2004), which confirms their conclusion on the presence of a significant amount of hidden magnetic energy and (unsigned) magnetic flux in the quiet solar photosphere.

\vspace{0.2truecm}

The reported patterns in the fractional linear polarization are of great diagnostic interest because they are sensitive to the thermal, dynamic and magnetic structure of the quiet solar atmosphere. We urge our solar observer colleagues to carry out the proposed observations as soon as possible. While a 1-m telescope with adaptive optics should be sufficient for the strongly polarizing Sr {\sc i} 4607 \AA\ line, the observation of this type of linear polarization signals in most of the other lines of the Second Solar Spectrum would require a larger aperture solar telescope\footnote{Note that the typical spatial and temporal scales of the reported $Q/I$ and $U/I$ fluctuations are those of the solar granulation.}. The design of the European Solar Telescope (EST) and of the Advanced Technology Solar Telescope (ATST) should be done thinking seriously in the scientific case of this Letter. To detect and quantify the predicted fluctuations in $Q/I$ and $U/I$ across a two-dimensional field of view would be of fundamental importance to decipher the small-scale magnetic activity of our nearest star. 

\acknowledgments
This research has been funded by the
Spanish Ministerio de Educaci\'on y Ciencia
through project AYA2004-05792 and by the National Academy of Sciences of Ukraine through project 1.4.6/7-226B.


\clearpage

\begin{figure}
\plotone{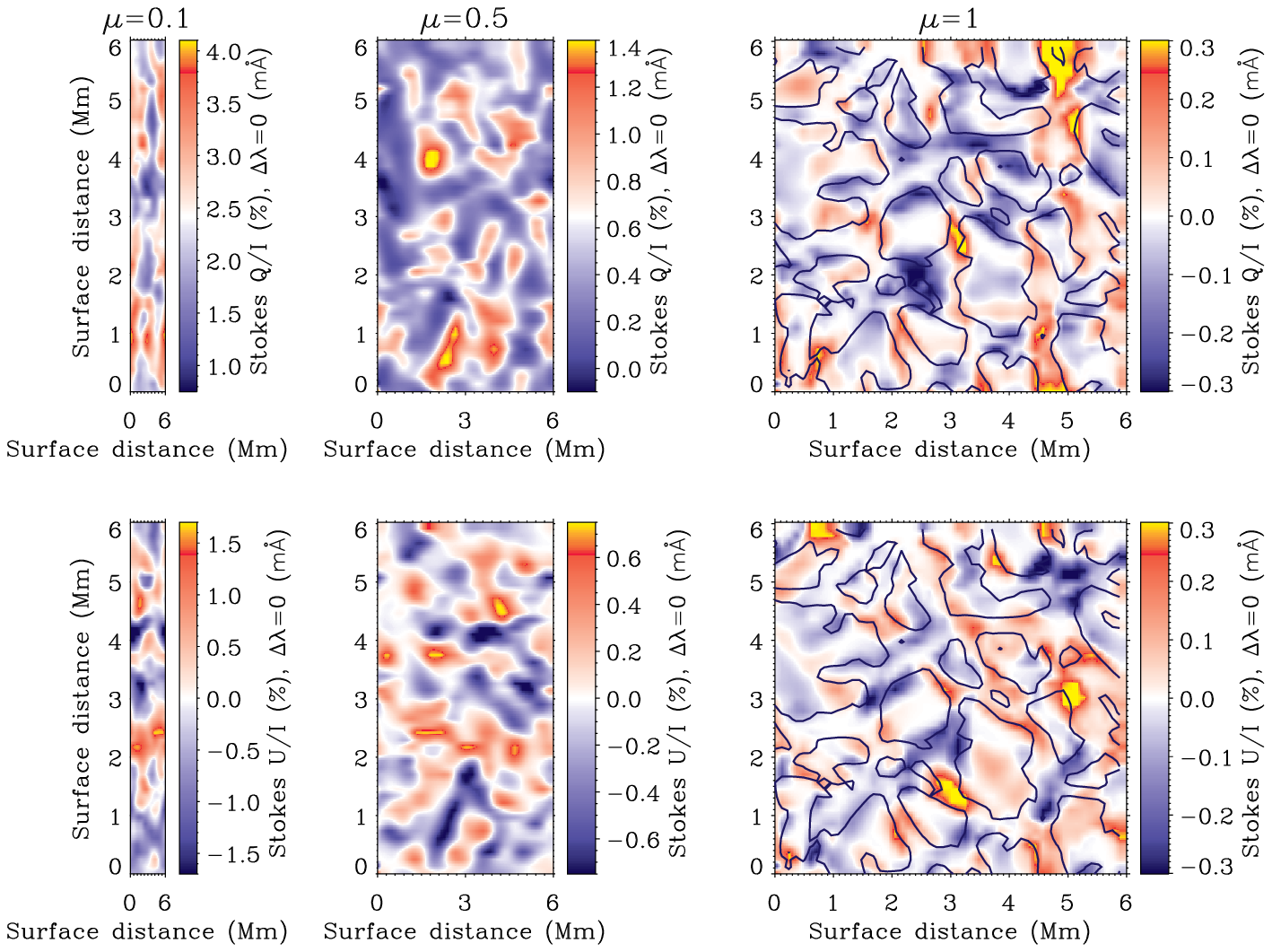}
\caption{The emergent $Q/I$ (top panels) and $U/I$ (bottom panels) at the line-center of the Sr {\sc i} 4607 \AA\ line calculated for three line-of-sights in a 3D snapshot of a realistic hydrodynamical simulation of solar surface convection and accounting for the diffraction limit effect of a 1-m telescope. The positive reference direction for Stokes $Q$ lies along the vertical direction of the corresponding panel, which for the $\mu=0.1$ and $\mu=0.5$ cases coincides with the parallel to the limb of the solar model. Note that we have taken into account the projection effects by means of which the off-disk-center images appear contracted by a factor $\mu$ along the 
horizontal direction of the figure panels. Note also that the ``surface distances" given in the plots measure the true separation between the points on the actual surface of the solar model. The solid-line contours in the $\mu=1$ panels delineate the (visible) upflowing granular regions.\label{fig_1}}
\end{figure}

\clearpage

\begin{figure}
\plotone{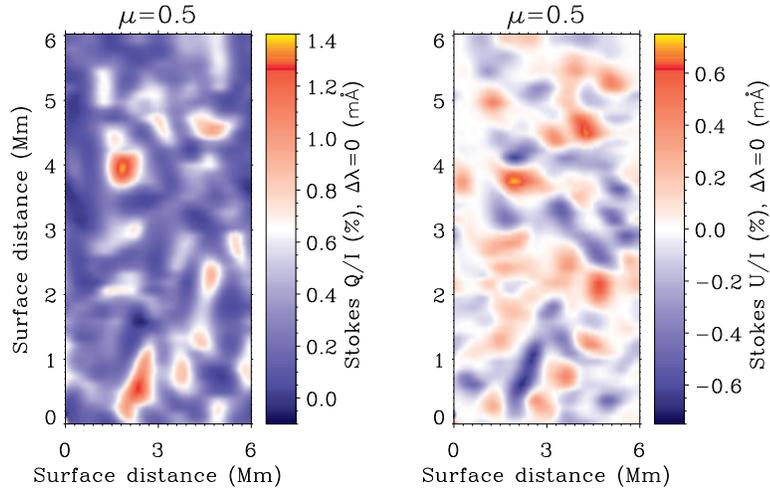}
\caption{The emergent $Q/I$ and $U/I$ signals at the line-center of the Sr {\sc i} 4607 \AA\ line calculated for a line-of-sight with $\mu=0.5$ 
in the 3D photospheric model, but taking into account the Hanle depolarization produced by the microturbulent magnetic field model discussed in the text. \label{fig_2}}
\end{figure}

\end{document}